\documentclass[twocolumn,showpacs,preprintnumbers,amsmath,amssymb,APSl,prd,nofootinbib,superscriptaddress]{revtex4-1}

\usepackage{dcolumn}
\usepackage{bm}
\usepackage{ifpdf}
\usepackage{hyperref}
\usepackage{bm}
\usepackage{xcolor,color,graphicx,graphics}
\usepackage[spanish,english]{babel}
\usepackage[latin1]{inputenc}
\usepackage[OT1]{fontenc}
\usepackage{latexsym,amssymb,amsmath,amsfonts}
\usepackage{makeidx}
\usepackage{epsfig,subfigure}
\usepackage{natbib}
\usepackage{epstopdf}
\usepackage{mathrsfs}
\usepackage{hyperref}
\hypersetup{colorlinks=true, linkcolor=blue, citecolor=green}
\usepackage{enumerate}

\definecolor{red}{rgb}{1,0,0}

\def\+{^\dagger}

\def\<{\leftarrow}
\def\>{\rightarrow}

\def\({\left(}
\def\){\right)}

\newcommand{\bi}{\begin{itemize}} 				\newcommand{\ei}{\end{itemize}}
\newcommand{\benu}{\begin{enumerate}} 		\newcommand{\enu}{\end{enumerate}}
\newcommand{\bd}{\begin{dinglist}{0}}     \newcommand{\ed}{\end{dinglist}}
\newcommand{\bfig}{\begin{figure}[htbp]}  \newcommand{\efig}{\end{figure}}
        			
\newcommand{\bc}{\begin{center}} 				  \newcommand{\ec}{\end{center}}
\newcommand{\be}{\begin{equation}} 				\newcommand{\ee}{\end{equation}}
\newcommand{\bsub}{\begin{subequations}}  \newcommand{\esub}{\end{subequations}}
\newcommand{\ben}{\begin{eqnarray}} 			\newcommand{\een}{\end{eqnarray}}
\newcommand{\ba}[1]{\begin{array}{#1}} 		\newcommand{\ea}{\end{array}}
\newcommand{\bea}{\begin{equation}\begin{array}{rcl}}
\newcommand{\eea}{\end{array}\end{equation}}

\begin{document}

\title{Junction conditions in Palatini $f(R)$ gravity}

\author{Gonzalo J. Olmo} \email{gonzalo.olmo@uv.es}
\affiliation{Departamento de F\'{i}sica Te\'{o}rica and IFIC, Centro Mixto Universidad de Valencia - CSIC.
Universidad de Valencia, Burjassot-46100, Valencia, Spain}
\affiliation{Departamento de F\'isica, Universidade Federal da
Para\'\i ba, 58051-900 Jo\~ao Pessoa, Para\'\i ba, Brazil}
\author{Diego Rubiera-Garcia} \email{drubiera@ucm.es}
\affiliation{Departamento de F\'isica Te\'orica and IPARCOS, Universidad Complutense de Madrid, E-28040 Madrid, Spain}

\date{\today}
\begin{abstract}

We work out the junction conditions for $f(R)$ gravity formulated in metric-affine (Palatini) spaces using a tensor distributional approach. These conditions are needed for building consistent models of gravitating bodies with an interior and exterior regions matched at some hypersurface. Some of these conditions depart from the standard Darmois-Israel ones of General Relativity and from their metric $f(R)$ counterparts. In particular, we find that the trace of the stress-energy momentum tensor in the bulk must be continuous  across the matching hypersurface, though its normal derivative need not to. We illustrate the relevance of these conditions by considering the properties of stellar surfaces in polytropic models, showing that the range of equations of state with potentially pathological effects is shifted beyond the domain of physical interest. This confirms, in particular, that neutron stars and white dwarfs can be safely modelled within the Palatini $f(R)$ framework. 

\end{abstract}

\maketitle

\section{Introduction}

Many models of gravitating bodies of both theoretical and observational interest involve the consideration of two different patches of space-time, glued at some hypersurface separating the interior from the exterior (see e.g. \cite{StephaniBook} and references therein). Such models can successfully yield both configurations with an interior filled with some matter distribution separated from an exterior vacuum solution, as in stellar bodies \cite{ComBook,Olmo:2019flu}, or two matter-filled regions such as in branes/domain walls \cite{RS1,RS2,Dvali:2000hr,Vilenkin:1984ib} or thin-shells \cite{Garcia:2011aa,Cardoso:2016oxy,Lobo:2020kxn}.
For these  constructions to be well defined from a mathematical point of view, a number of conditions must hold at the matching hypersurface, which within the context of General Relativity (GR) are known as the Darmois-Israel matching (or junction) conditions \cite{Darmois,Israel:1966rt}. These conditions involve the continuity of the first fundamental form across the matching hypersurface, and a number of conditions relating the stress-energy tensor with the discontinuity in the extrinsic curvature on that hypersurface.

Given the physical relevance of such models and conditions for the modeling of different bodies, a timely question is to what extent they can be applied to other theories of gravity beyond GR (see e.g. \cite{CLreview,Clifton:2011jh,Nojiri:2017ncd,Heisenberg:2018vsk} for some reviews). The consideration of these theories are motivated by a variety of reasons in view of the challenges posed by the exploration of the strong-field regime of the gravitational interaction in the multimessenger astronomy era \cite{Berti:2015itd}, and by the difficulties of the early and late-time cosmological models \cite{Bull:2015stt}. Typically such theories introduce additional difficulties in the form of higher-order equations of motion and much more involved dynamics, which naturally requires an upgrade of the junction conditions \cite{Davis:2002gn,Gravanis:2002wy,Deruelle:2007pt,Padilla:2012ze,Vignolo:2018eco,delaCruz-Dombriz:2014zaa}. This introduces additional constraints, which troubles the construction of such matched models.  Perhaps the simplest family of extensions of GR are those based on a function of the curvature scalar, the so-called $f(R)$ gravity, which has attracted a great deal of attention in the literature \cite{DeFelice:2010aj}. In this case, it has been proven that many known matched solutions in GR do not survive when considered in the $f(R)$ scenario \cite{Senovilla:2013vra}. Thus, the analysis of the junction conditions beyond GR is of utmost importance in order to consistently explore the phenomenology of such models.

The main aim of this paper is to work out the junction conditions corresponding to the Palatini formulation of $f(R)$ theories. In the Palatini approach, metric and affine connection are regarded as independent degrees of freedom, and the field equations are obtained from independent variations of the action with respect to both of them \cite{Olmo:2011uz}. As opposed to their metric counterpart, the resulting theory has second-order field equations, with the new gravitational dynamics being due to new nonlinear couplings engendered by the matter fields, and no extra propagating degrees of freedom are introduced. Moreover, Palatini $f(R)$ models may pass the post-Newtonian (solar system) tests \cite{Olmo:2005hc}, are compatible with current gravitational wave observations, and have a rich phenomenology in black hole \cite{Bambi:2015zch}, stellar \cite{Olmo:2019qsj}, and cosmological \cite{Enckell:2018hmo} scenarios.

The equations of motion and junction conditions corresponding to Palatini $f(R)$ theories turn out to be clearly different from those corresponding to GR and to the metric formulation of $f(R)$ theories. Differences arise even in the conservation equations on the matching hypersurface, being the same in GR and in metric $f(R)$ \cite{Senovilla:2013vra} but different in Palatini $f(R)$ gravity.
Our results will also prove useful, in particular, to better understand an issue that has undermined progress in the study of stellar models in Palatini $f(R)$ theories in the last years. We are referring to the potential emergence of curvature divergences near the surface of polytropic stellar models \cite{Olmo:2008pv,Barausse:2007pn,Barausse:2007ys,Barausse:2008nm}.  As we will see, a consistent distributional treatment of the field equations puts forward that such divergences are an artifact of the approach followed in \cite{Barausse:2007pn,Barausse:2007ys,Barausse:2008nm,Olmo:2008pv} and that the equations on the matching surface are free from such pathologies for polytropic equations of state of physical interest.


\section{Junction conditions for $f(R)$ gravity in Palatini approach}

\subsection{Action and field equations}

The action of $f(R)$ gravity can be written as
\begin{equation} \label{eq:actionfR}
\mathcal{S}=\frac{1}{2\kappa^2} \int d^4 x \sqrt{-g} f(R) + \int d^4x \sqrt{-g} \mathcal{L}_m(g_{\mu\nu},\psi_m)  \ ,
\end{equation}
where $\kappa^2$ is Newton's constant in suitable units, $g$ is the determinant of the space-time metric $g_{\mu\nu}$, $f(R)$ is some function of the curvature scalar $R \equiv g^{\mu\nu}R_{\mu\nu}(\Gamma)$, with $\Gamma \equiv \Gamma_{\mu\nu}^{\lambda}$ the affine connection, and $\mathcal{L}_m$ denotes the matter Lagrangian with $\psi_m$ representing a collection of matter fields.

In the metric formalism the connection is taken to be given by the Levi-Civita one, that is, the one satisfying $\nabla_{\mu}^{\Gamma}(\sqrt{-g} g^{\alpha\beta})=0$, and the variation of the action (\ref{eq:actionfR}) with respect to the metric yields the field equations
\begin{eqnarray}\label{eq:fRmet}
f_R G_{\mu\nu}(g)&-&\frac{Rf_R-f}{2}g_{\mu\nu} -f_{RR}(\nabla_{\mu}\nabla_{\nu} R -g_{\mu\nu}\nabla_{\alpha}\nabla^{\alpha}R) \nonumber \\
&-&f_{RRR}(\nabla_{\mu}R\nabla_{\nu} R -g_{\mu\nu}\nabla^{\alpha}\nabla_{\alpha}R)=\kappa^2 T_{\mu\nu} \ , 
\end{eqnarray}
where $f_R \equiv df/dR$, and $T_{\mu\nu}=\frac{2}{\sqrt{-g}} \frac{\delta \mathcal{S}_m}{\delta g^{\mu\nu}}$ is the stress-energy tensor. Tracing in this equation with $g^{\mu\nu}$ yields
\begin{equation}\label{eq:tracemet}
3\Box f_R + Rf_R - 2f=\kappa^2 T \ ,
\end{equation}
where $T$ is the trace of the stress-energy tensor. As is well known, the presence of the derivative operator $\Box$ yields a dynamical character to the (scalar) field $\phi \equiv f_R$, which can be read off as an additional propagating degree of freedom.

On the other hand, in the Palatini formulation, the field equations follow by independent variations of the action (\ref{eq:actionfR}) with respect to the metric and the connection. After solving the connection equation, the resulting equations for the metric can be conveniently cast as \cite{Sotiriou:2008rp,Olmo:2011uz}
\begin{eqnarray} \label{eqs:fR}
f_R G_{\mu\nu}(g)&+&\frac{Rf_R-f}{2}g_{\mu\nu} - [\nabla_{\mu}\nabla_{\nu}f_R-g_{\mu\nu} \Box f_R] \\
&+&\frac{3}{2f_R}[\nabla_{\mu}f_R \nabla_{\nu}f_R - \frac{1}{2} g_{\mu\nu}(\partial f_R)^2]=\kappa^2 T_{\mu\nu} \ , \nonumber
\end{eqnarray}
with a similar notation as in the metric case. Tracing (\ref{eqs:fR}) with $g^{\mu\nu}$ yields
\begin{equation} \label{eq:tracepal}
Rf_R-2f=\kappa^2 T \ .
\end{equation}
This equation implies that $R\equiv R(T)$ which, instead of a differential equation, is simply an algebraic relation linking curvature and matter. This fact introduces a radical difference in the properties and dynamics of the theory with respect to the metric formulation and, therefore, a distributional analysis of the Palatini $f(R)$ equations must take care of that fact.

Let us point out that the above equations (\ref{eq:fRmet}) and (\ref{eqs:fR}) for the metric and the Palatini versions of $f(R)$ theories can be seen as particular cases of Brans-Dicke scalar-tensor theories with a potential. In particular, the trace equations (\ref{eq:tracemet}) and (\ref{eq:tracepal}) can be seen as specific cases of the general equation
\begin{equation} \label{eq:traceBD}
(3+2\omega)\Box\phi+2V(\phi)-\phi V_\phi=\kappa^2T \ ,
\end{equation}
with the metric case corresponding to $\omega=0$ while Palatini having $\omega=-3/2$. The scalar field is identified as $\phi\equiv f_R$ and  its potential is given by $V(\phi)=\phi R(\phi)-f(R(\phi))$. Note that the scalar field is non-dynamical in the Palatini case, while in any other case it does propagate. In the following, we will consider the distributional description of Eq.(\ref{eqs:fR}), which is equivalent to the $\omega=-3/2$ Brans-Dicke theory.

\subsection{Distributional stress-energy tensor}

We consider two smooth four-dimensional manifolds $(M^{\pm},g_{\mu\nu})$, and let us denote by $V^{\pm}$ two bounded regions living in $M^{\pm}$ with boundaries $\Sigma^{\pm}$. These regions are matched (\textit{``glued"}) at a time-like hypersurface $\Sigma$ with the natural identification of their boundaries as $\Sigma^{+}=\Sigma^{-}$, and with the space-time metric $g_{\mu\nu}$ assumed to be well defined through the entire manifold and, in particular, to be continuous (but not necessarily differentiable) at $\Sigma$, which is a basic requirement in the junction conditions formalism \cite{ClaDra}. Since there may be discontinuities in several geometric quantities across $\Sigma$, the suitable tool to deal with this scenario is that of tensorial distributions, namely, tensor fields with compact support on the manifold. One of the main goals of the junction conditions is to identify the singular parts of the curvature and matter tensorial distributions on $\Sigma$ and find the allowed discontinuities of these quantities across it. For the sake of our analysis we borrow, to a large extent, the notation introduced in Ref.\cite{Mars:1993mj}, and we refer the reader there for further details on the mathematical setup employed here.

Given the particular way the matter fields feed the new dynamics in Palatini $f(R)$ gravity, as follows from Eq.(\ref{eq:tracepal}), we start our analysis by introducing a distribution for the stress-energy tensor as
\begin{equation} \label{eq:Tmunu}
\underline{T}_{\mu\nu}=T_{\mu\nu}^{+} \underline{\theta} +T_{\mu\nu}^{-}(\underline{1}-\underline{\theta}) + \tau_{\mu\nu}\underline{\delta}^{\Sigma} \ .
\end{equation}
Hereafter underbars indicate distributions; $T_{\mu\nu}^{\pm}$ are the stress-energy tensors in $V^{\pm}$, respectively; $\underline{\theta}$  is the scalar distribution defined by the (locally integrable) Heaviside function, the latter taking the value of $1$ in $V^+$, $0$ in $V^-$, and any intermediate reference value in $\Sigma$; $\underline{\delta}^{\Sigma}$ is a scalar Delta-type distribution with support on $\Sigma$ acting upon any test function $X$ as $<\underline{\delta},X>\equiv\int_{\Sigma} X$; and $\tau_{\mu\nu}$ accounts for the singular part of the stress-energy tensor on $\Sigma$. Similarly, the distributional form of the trace of the stress-energy tensor reads
\begin{equation} \label{eq:T}
\underline{T}=T^{+} \underline{\theta} +T^{-}(\underline{1}-\underline{\theta}) + \tau \underline{\delta}^{\Sigma} \ ,
\end{equation}
where $\tau \equiv {\tau^\mu}_{\mu}$ is the trace of the singular part of the stress-energy tensor.

To cast the field equations (\ref{eqs:fR}) in distributional form, we have to deal with several potential difficulties in the transition from regular tensorial functions to tensorial distributions.  Let us isolate first the term
\begin{equation} \label{eq:nabfr}
\nabla_{\mu}f_R \nabla_{\nu} f_R =f_{RR}^2 R_T^2 \nabla_{\mu} T\nabla_{\nu}T \ ,
\end{equation}
where $R_T \equiv dR/dT=\kappa^2/(Rf_{RR}-f_R)$. From (\ref{eq:T}), the distributional form of the covariant derivative of the trace of the stress-energy tensor  reads
\begin{equation}
\nabla_{\mu} \underline{T} = \nabla_{\mu} T^+ \underline{\theta} + \nabla_{\mu} T^- (\underline{1}-\underline{\theta}) + n_{\mu} [T] \underline{\delta}^{\Sigma} +\nabla_{\mu} (\tau \underline{\delta}^{\Sigma}) \ .
\end{equation}
Here $n_{\mu}$ is the unit vector normal to $\Sigma$ defined via $\nabla_{\mu} \underline{\theta}=n_{\mu} \underline{\delta}^\Sigma$, and brackets represent a discontinuity (``jumps") across $\Sigma$ in the quantity contained there. Thus $[T] \equiv T^+-T^-$ is the jump in the trace of the stress-energy tensor. Given that the term $\nabla_{\mu} T\nabla_{\nu}T$ in (\ref{eq:nabfr}) leads to products of the form $\underline{\theta} \cdot \underline{\delta}$ and $\underline{\delta} \cdot \underline{\delta}$, which are not well defined in distributional theory, in order to have consistent equations in distributional sense we must require the conditions
\begin{eqnarray} \label{eq:jumpbrane}
[T]&=&0 \\
\tau&=&0 \label{eq:jumpbrane2} \ .
\end{eqnarray}
The first of these conditions simply indicates the need for the continuity of the trace of the stress-energy tensor across $\Sigma$, which was already expected on grounds of continuity and standard differentiability of the tensorial equations. The second condition, $\tau=0$, is new and entirely due to the distributional nature of the quantities defined on the hypersurface. In the context of GR, it would imply a vanishing brane tension. In the modified gravity context, its implications are still to be understood. Note that these restrictions involve the traces of the stress-energy tensor and its singular part on $\Sigma$, but tell us nothing about the full structure of these objects.

Bearing in mind (\ref{eq:jumpbrane}), let us consider next the contribution to the field equations (\ref{eqs:fR}) given by the term (in distributional sense)
\begin{equation}
\nabla_{\mu}\nabla_{\nu} \underline{T}=\nabla_{\mu}\nabla_{\nu} T^+ \underline{\theta} + \nabla_{\mu} \nabla_{\nu} T^- (\underline{1}-\underline{\theta})+ n_{\mu}n_{\nu} b \underline{\delta}^{\Sigma} \ ,
\end{equation}
where we have defined the scalar quantity
\begin{equation}
b \equiv n^{\alpha} [\nabla_{\alpha} T] \ ,
\end{equation}
which is the discontinuity in the covariant derivative of the trace in the normal direction to the hypersurface $\Sigma$. This expression shall be needed later.

\subsection{Decomposition of the Einstein tensor on $\Sigma$}

The distributional form of the Einstein tensor takes the form \cite{ClaDra}
\begin{equation}
\underline{G}_{\mu\nu}=G_{\mu\nu}^+ \underline{\theta} +G_{\mu\nu}^- (\underline{1}-\underline{\theta}) + \mathcal{G}_{\mu\nu} \underline{\delta}^{\Sigma} \ ,
\end{equation}
where $\mathcal{G}_{\mu\nu}$ represents the singular part of the Einstein tensor on $\Sigma$. Plugging this expression in the field equations  (\ref{eqs:fR}), and taking into consideration the constraint (\ref{eq:jumpbrane}) on the stress-energy tensor, the singular part of the field equations (\ref{eqs:fR}) can be written as
\begin{equation} \label{eq:singG}
f_{R_\Sigma} \mathcal{G}_{\mu\nu}+  \left. f_{RR} R_T  \right|_\Sigma b h_{\mu\nu}=\kappa^2  \tau_{\mu\nu} \ ,
\end{equation}
where the subindex $\Sigma$ denotes quantities evaluated on $\Sigma$, while $h_{\mu\nu}$ is the projector on $\Sigma$, namely
\begin{equation} \label{eq:firstform}
h_{\mu\nu}\equiv g_{\mu\nu} -n_{\mu}n_{\nu} \ ,
\end{equation}
usually referred to as the first fundamental form, and which must be continuous across $\Sigma$ \cite{ClaDra}. Given that  $\mathcal{G}_{\mu\nu}$ can be expressed as \cite{Mars:1993mj}
\begin{equation} \label{eq:singpart}
\mathcal{G}_{\mu\nu}=-[K_{\mu\nu}]+h_{\mu\nu} [K^{\rho}_{\rho}] \ ,
\end{equation}
where
\begin{equation}
K_{\mu\nu}^{\pm} \equiv {h^\rho}_{\beta} {h^\sigma}_{\mu} \nabla_{\rho}^{\pm} n_{\sigma} \ ,
\end{equation}
is the second fundamental form on $V^{\pm}$, respectively, we can trace in Eq.(\ref{eq:singG}) with the metric $g^{\mu\nu}$ and using (\ref{eq:jumpbrane2}) together with the fact that (\ref{eq:singpart}) implies that $\mathcal{G}_{\rho}^{\rho}=2[K_{\rho}^{\rho}]$, one arrives to
\begin{equation} \label{eq:Krhorho}
[K^{\rho}_{\rho}] =-\left.\frac{3f_{RR}R_T}{2f_R} \right|_\Sigma b \ .
\end{equation}
Plugging this relation back in (\ref{eq:singG}) one finally gets
\begin{equation} \label{eq:Kfinal}
-[K_{\mu\nu}]+\frac{1}{3}h_{\mu\nu}[K^{\rho}_{\rho}]=\kappa^2 \frac{\tau_{\mu\nu}}{f_{R_\Sigma}} \ ,
\end{equation}
which is clearly consistent with the tracelessness of $\tau_{\mu\nu}$ in (\ref{eq:jumpbrane2}).
Note that these results largely depart from the corresponding expressions in GR, where the singular part of the Einstein equations $\mathcal{G}_{\mu\nu}=\kappa^2 \tau_{\mu\nu}$ does not introduce the constraint (\ref{eq:jumpbrane2}), since the latter is associated to the contribution (\ref{eq:nabfr}), which is vanishing when $f(R)=R-2\Lambda$ (the Einstein-Hilbert action with cosmological constant term $\Lambda$). Thus, in such a case, instead of (\ref{eq:Kfinal}) one has $-[K_{\mu\nu}]+h_{\mu\nu}[K^{\rho}_{\rho}]=\kappa^2 \tau_{\mu\nu}$, whose trace reads $2[K^{\rho}_{\rho}]=\kappa^2 \tau$ and, therefore, the brane tension in GR is non-vanishing in general. As opposed to that, we have just seen that in Palatini $f(R)$ gravity the brane tension vanishes but $[K^{\rho}_{\rho}] \neq 0$. In the metric version of $f(R)$, however, one typically  has $[K^{\rho}_{\rho}]=0$ \cite{Senovilla:2013vra}. It is important to note that the condition (\ref{eq:jumpbrane2}) imposes a constraint between the energy density and the two principal pressures on $\Sigma$, which effectively reduces the number of degrees of freedom of the matter on the hypersurface $\Sigma$. This is expected to have important implications for the resolution of the conservation equations.

\subsection{Energy conservation equations}

As shown in \cite{Mars:1993mj}, the  Bianchi identities hold in the distributional sense, that is $\nabla_\mu\underline{G}^\mu_{\ \nu}=0$. This can be explicitly written as two sets of equations:
\begin{eqnarray}
(K_{\rho\sigma}^{+} + K_{\rho\sigma}^-) \mathcal{G}^{\rho\sigma}&=&2n^{\rho}n^{\sigma}[R_{\rho\sigma}]-[R] \label{eq:ec1} \\
D^{\rho}\mathcal{G}_{\rho\nu}&=&-n^{\rho}h^{\sigma}_{\nu}[R_{\rho\sigma}] \ ,  \label{eq:ec2}
\end{eqnarray}
where $D_{\rho}\equiv {h_\rho}^{\alpha} \nabla_{\alpha}$ denotes the covariant derivative on $\Sigma$. On the other hand, given that (\ref{eq:tracepal}) leads to $R=R(T)$ and that $[T]=0$, it follows that $[R]=0$. Our next goal is to extract additional information from the two equations (\ref{eq:ec1}) and (\ref{eq:ec2}) about the behavior of the stress-energy tensor on the hypersurface $\Sigma$. To this end, let us start by taking a covariant derivative upon Eq.(\ref{eq:singG}) to find
\begin{eqnarray} \label{eq:Drho}
\kappa^2 D_{\rho} {\tau^\rho}_{\nu}&=& \left. f_{RR} R_T  \right|_\Sigma ((D_{\rho} T_{\Sigma}){\mathcal{G}^\rho}_{\nu}+D_{\nu} b) +f_{R_\Sigma} D_{\rho} {\mathcal{G}^\rho}_{\nu} \ ,  \nonumber \\
&+&\left. (f_{RRR} R_T^2+ f_{RR}R_{TT})\right|_\Sigma b(D_{\nu}T_{\Sigma}) \ ,
\end{eqnarray}
where we have used the relation $D_{\nu} f_R=  \left. f_{RR} R_T  \right|_\Sigma(D_{\nu} T_{\Sigma})$ with $T_{\Sigma} $ denoting the value of $T$ on $\Sigma$. Let us deal with the different terms in this expression, starting with the one in $[R_{\mu\nu}]$, as follows from the replacement of (\ref{eq:ec2}) in (\ref{eq:Drho}). Projecting on the field equations (\ref{eqs:fR}) with $n^{\rho}{h^\sigma}_{\nu}$ we find
\begin{eqnarray}
&&f_R n^{\rho}{h^\sigma}_{\nu} R_{\rho\sigma} - n^{\rho}{h^\sigma}_{\nu}\nabla_{\rho}\nabla_{\sigma} f_R \nonumber \\
&+&\frac{3}{2f_R}(n^{\rho}\nabla_{\rho}f_R)f_{RR}R_T (D_{\nu}T_{\sigma})=\kappa^2 n^{\rho}{h^\sigma}_{\nu} T_{\rho\sigma} \ .
\end{eqnarray}
The jump of this expression across $\Sigma$ becomes
\begin{eqnarray} \label{eq:Rrhosig}
f_R n^{\rho} {h^\sigma}_{\nu}[R_{\rho\sigma}]&=&\kappa^2 n^{\rho} {h^\sigma}_{\nu}[T_{\rho\sigma}] + n^{\rho}{h^\sigma}_{\nu}[\nabla_{\rho}\nabla_{\sigma} f_R] \nonumber \\
&-&\frac{3f_{RR}^2}{2f_R} R_T^2 b (D_{\nu} T_{\Sigma}) \ .
\end{eqnarray}
Bearing in mind that
\begin{eqnarray}
&&\nabla_{\rho} \nabla_{\sigma} f_R =\nabla_{\rho}(f_{RR} R_T \nabla_{\sigma} T) \\
&=&(f_{RRR}R_T^2+f_{RR}R_{TT})\nabla_{\rho}T\nabla_{\sigma}T + f_{RR} R_T\nabla_{\rho}\nabla_{\sigma} T \ , \nonumber
\end{eqnarray}
then its jump across $\Sigma$ can be written as
\begin{eqnarray} \label{eq:intste}
n^{\rho}{h^\sigma}_{\nu}[\nabla_{\rho}\nabla_{\sigma}f_R]&=&(f_{RRR}R_T^2 + f_{RR}R_{TT})b(D_{\nu}T_{\Sigma}) \nonumber \\
&+&f_{RR}R_T n^{\rho}{h^\sigma}_{\nu}[\nabla_{\rho}\nabla_{\sigma} T] \ .
\end{eqnarray}
To deal with the last term in this expression we introduce, following \cite{Senovilla:2013vra}, a decomposition of the form
\begin{eqnarray}
[\nabla_{\rho}\nabla_{\sigma} T]&=&An_{\rho}n_{\sigma}+n_{\rho}((D_{\sigma} b) -[K_{\rho\sigma}](D^{\rho} T_{\Sigma})) \\
&+&n_{\sigma}((D_{\rho}b)-[K_{\rho\nu}](D^{\rho} T_{\Sigma}))+\frac{b}{2}(K_{\rho\sigma}^{+}+K_{\rho\sigma}^{-}) \ , \nonumber
\end{eqnarray}
where $A \equiv n^{\mu}n^{\nu}[\nabla_{\mu}\nabla_{\nu}R]$. This way, Eq.(\ref{eq:intste}) reads
\begin{equation} \label{eq:hsig}
n^{\rho}{h^\sigma}_{\nu}[\nabla_{\rho}\nabla_{\sigma} T]=(D_{\nu}b) - [K_{\alpha\nu}](D^{\alpha} T_{\Sigma}) \ .
\end{equation}
We are almost there. Combining Eqs.(\ref{eq:Rrhosig}), (\ref{eq:intste}) and (\ref{eq:hsig}) with (\ref{eq:Drho}), and using (\ref{eq:ec2}) and (\ref{eq:Krhorho}) we find, after a bit of algebra, that (\ref{eq:Drho}) simplifies down to
\begin{equation} \label{eq:main2}
D^{\rho}\tau_{\rho\nu}=-n^{\rho}{h^\sigma}_{\nu} [T_{\rho\sigma}] \ ,
\end{equation}
which coincides with the expression found  in GR and in metric $f(R)$ gravity.

Let us now extract information from Eq.(\ref{eq:ec1}). The most important point here is to compute the quantity on its left-hand side using the field equations (\ref{eqs:fR}). Projecting with $f_R n^{\rho}n^{\sigma}$ and comparing the result on both sides of $\Sigma$ recalling that $g_{\mu\nu}$ must be continuous across $\Sigma$, one finds
\begin{eqnarray} \label{eq:ec2m}
f_R n^{\rho}n^{\sigma}[G_{\rho\nu}] &+& h^{\rho\sigma}[\nabla_{\rho}\nabla_{\sigma}f_R] - \frac{3}{2f_R} h^{\rho\sigma}[\nabla_{\rho}f_R \nabla_{\sigma} f_R] \nonumber \\
&+&\frac{3}{4f_R}[(\nabla_{\mu} f_R)^2]=\kappa^2 n^{\rho}n^{\sigma}[T_{\rho\sigma}] \ .
\end{eqnarray}
Let us isolate the various contributions to this equation. First we have the term
\begin{eqnarray}
h^{\rho\sigma}[\nabla_{\rho}\nabla_{\sigma} f_R] &=& (f_{RRR}R_T^2+f_{RR}R_{TT})[(D_{\rho} T_{\Sigma})^2] \nonumber \\
&+& f_{RR}R_Th^{\rho\sigma}[\nabla_{\rho}\nabla_{\sigma}T] \nonumber \\
&=&\frac{b}{2} f_{RR}R_T h^{\rho\sigma}(K_{\rho\sigma}^+ + K_{\rho\sigma}^-) \ ,
\end{eqnarray}
where in the last line we have explicitly implemented the fact that, since $T$ must  be continuous across $\Sigma$, then  $(D_{\rho} T_{\Sigma})=0$.  The second piece of (\ref{eq:ec2m}) that we need to consider is
\begin{eqnarray}
h^{\rho\sigma}[\nabla_{\rho}f_R \nabla_{\sigma} f_R]&=&f_{RR}^2R_T^2h^{\rho\sigma}[\nabla_{\rho} T\nabla_{\sigma} T] \nonumber \\
&=&f_{RR}^2 R_T^2 [(D_{\rho} T_{\Sigma})^2]=0 \ ,
\end{eqnarray}
the last equality due to the same considerations as before. The third relevant piece of (\ref{eq:ec2m}) is
\begin{equation}
[(\nabla_{\mu} f_R)^2]=(n^{\mu}n^{\nu}+h^{\mu\nu})\nabla_{\mu}f_R \nabla_{\nu}f_R=R_T^2 f_{RR}^2 [b^2] \ ,
\end{equation}
where we have defined $[b^2] \equiv [(n^{\alpha}\nabla_{\alpha}T)^2]=2b n^\alpha \partial_\alpha T_\Sigma$ (see Appendix D.2. of Ref.\cite{Reina:2015gxa} for details). With all these elements, Eq.(\ref{eq:ec2m}) reads
\begin{eqnarray}
f_R n^{\rho}n^{\sigma}[G_{\rho\nu}]&=&\kappa^2 n^{\rho}n^{\sigma}[T_{\rho\sigma}] -f_{RR}R_T h^{\rho\sigma}\frac{b}{2}(K_{\rho\sigma}^+ + K_{\rho\sigma}^- ) \nonumber \\
&-&\frac{3R_T^2 f_{RR}^2}{f_R}[b^2] \ .
\end{eqnarray}
Plugging this expression in the right-hand-side of Eq.(\ref{eq:ec2m}) we finally find
\begin{equation} \label{eq:main3}
(K_{\rho\sigma}^+ + K_{\rho\sigma}^-)\tau^{\rho\sigma} =2n^{\rho}n^{\sigma}[T_{\rho\sigma}] -\frac{3R_T^2 f_{RR}^2}{f_R} [b^2] \ .
\end{equation}
Likewise the analysis of the second fundamental form above, this result recovers the one of GR only if $f_{RR}=0$, that is, $f(R)=R-2\Lambda$.

To summarize the main results of this section, the junction conditions in Palatini $f(R)$ gravity are: the continuity of the first fundamental form (\ref{eq:firstform}) across $\Sigma$; the continuity of the trace of the stress-energy tensor across $\Sigma$, namely Eq.(\ref{eq:jumpbrane}); the  vanishing of the trace of its singular part on $\Sigma$, namely Eq.(\ref{eq:jumpbrane2}); and Eq.(\ref{eq:Kfinal}) for the discontinuity of the trace of the second fundamental form, which we recall it departs from the GR and the metric $f(R)$ cases \cite{Senovilla:2013vra}. In cases allowing for shells or branes, the energy content on $\Sigma$ is determined by Eqs.(\ref{eq:Kfinal}), (\ref{eq:main2}) and (\ref{eq:main3}).

\section{Application: polytropic stellar models}

For a specific implementation of this framework, we will now apply the formulas derived in the previous section to address an issue raised in \cite{Barausse:2007pn,Barausse:2007ys,Barausse:2008nm} that affects models of stellar structure in Palatini $f(R)$ gravity. Using the standard tensorial approach, it turns out that when one attempts to match a polytropic stellar model with equation of state
\begin{equation} \label{eq:poly}
\rho(P)=\left(\frac{P}{K}\right)^{1/\gamma}+\frac{P}{\gamma-1} \ ,
\end{equation}
to an external Schwarzschild solution, certain difficulties arise. Curvature divergences, in particular,  develop at the matching surface, the latter characterized by vanishing pressure. Knowing that this happens for a range of polytropic indices of physical interest, namely, $3/2<\gamma<2$, which includes the relevant case $\gamma=5/3$ that describes a gas of non-relativistic degenerate fermions, one might be tempted to conclude that the viability of Palatini $f(R)$ models could be in danger. Yet one should bear in mind that polytropic models are just crude statistical approximations of some stars. On the other hand, given that the divergent terms are directly related to what happens at the matching hypersurface, it seems appropriate to use the more rigorous approach of tensorial distributions developed above to reassess the problem. Under this light, it seems relevant to see what Eq.(\ref{eq:singG}), which describes the dynamics at the matching hypersurface, has to say about the behavior of curvature there.

A glance at Eq.(\ref{eq:singG}) puts forward that the discontinuity in the derivative of the trace of the stress-energy tensor in the direction normal to $\Sigma$ is the key element that controls the modified dynamics on $\Sigma$. In fact, given that the trace $T$ must be continuous across the surface (as follows from (\ref{eq:jumpbrane})) and that $R=R(T)$ must also be (due to (\ref{eq:tracepal})) the coefficient $f_{R_\Sigma}$ must be smooth and generically expected to be close to unity in regions of low energy density such as the surface of a star. This is the case, for instance, of the quadratic model $f(R)=R+a R^2$, for which Eq.(\ref{eq:tracepal}) yields $R(T)=-\kappa^2 T$ and we have $f_R=1$ as $T\to 0$. Thus, whenever $b \equiv n^{\alpha} [\nabla_{\alpha} T]$ is not zero, the nonlinearities of the $f(R)$ theory will contribute to the surface dynamics. Continuing with the quadratic model, it is easy to see that the term $f_{RR} R_T=-2 a\kappa^2$ is just a constant and, therefore, $b$ is the element that may bring in new effects. Considering a polytrope with equation of state (\ref{eq:poly}) it follows that $b=n^r_+\partial_r T^+-n^r_-\partial_rT^-=-n^r_-\partial_r T^-\approx (3-\rho_P)P_r$. According to the analysis of \cite{Olmo:2008pv}, one finds that $\rho_P\sim P^{(1-\gamma)/\gamma}$ and $P_r\sim P^{1/\gamma}$, which leads to $b\sim  P^{(2-\gamma)/\gamma}$. It turns out that this quantity only diverges if $\gamma>2$, shifting the problematic range of $\gamma$ beyond the domain of direct physical interest. For $\gamma<2$, it is evident that $b=0$ at the surface. This makes Eq.(\ref{eq:singG}) formally identical to its GR counterpart, as one would expect in any model which only departs from GR at high curvature/densities. The conservation equations (\ref{eq:main2}) and (\ref{eq:main3}) also degenerate into their GR counterparts.

We have just seen that a careful analysis of the junction conditions using distribution theory has been able to get rid of a problem that arises when a standard tensorial approach is considered. In the usual approach, in order to have a well defined geometry, the trace of the matter stress-energy tensor must be continuous and differentiable because, otherwise, the second derivatives of the matter fields that appear in the field equations would lead to divergences. In the distributional approach, however, we have seen that those requirements are slightly relaxed at a matching hypersurface because one is forced to have a continuous $T$ but a normal discontinuity in $\nabla_\mu T$ is allowed and still provides sensible results. A glance at the analysis of \cite{Olmo:2008pv} shows that the divergent terms that arise at the surface of polytropic models are entirely due to the evaluation of the term $\nabla_\mu\nabla_\nu T$ at the matching surface, where $P=0$. Specifically, $\nabla_\mu\nabla_\nu T$  generates two\footnote{We note  that the divergence due to $\rho_P P_{rr}$ was overlooked in \cite{Olmo:2008pv}.} terms  of the form $\rho_P P_{rr}$ and $\rho_{PP}P_r^2$ which go as $\sim  P^{(3-2\gamma)/\gamma}$ and diverge for $\gamma>3/2$ when $P\to 0$.
In the distributional approach, however, the terms with $\nabla_\mu\nabla_\nu T$ provide a contribution on $\Sigma$ proportional to $b$, which measures the discontinuity in $\nabla_\mu T$. This property is precisely what cures the singular terms on the matching surface, since the seemingly pathological contributions $\rho_P P_{rr}$ and $\rho_{PP}P_r^2$ are now replaced by  $\rho_P P_r$. Thus, a smooth matching surface between an external Schwarzschild solution and an internal polytropic fluid in Palatini $f(R)$ gravity is possible whenever $\gamma<2$ when a proper distributional treatment of the matching surface is considered.

\section{Conclusion}

In this work we have derived the junction conditions corresponding to Palatini $f(R)$ theories using distribution theory. We have found that the equations that relate the extrinsic curvature to the matter sources on the matching surface differ from those found in GR and in the metric version of $f(R)$ theories. This is a non-trivial result which has many repercussions for the building of matched solutions of any kind within these theories, and which must be carefully implemented in order not to get to misleading results.

As an specific example, we have considered the relevant case of polytropic stellar models. In such a case, a naive approach to match the internal fluid solution to the external Schwarzschild solution, as followed in \cite{Olmo:2008pv,Barausse:2007pn,Barausse:2007ys,Barausse:2008nm}, leads to the unacceptable emergence of curvature divergences at the matching surface for a range of polytropic indices of physical interest. Here we have shown that a more rigorous treatment using distribution theory modifies this conclusion, shifting the range of potentially pathological equations of state beyond the domain of relevance for physical applications. The junction conditions derived here, therefore, should be applied whenever one attempts to build stellar models that match vacuum external solutions in any Palatini $f(R)$ theory allowing, in particular, to consistently model neutron stars and white dwarfs, among others.

Further applications of these junctions conditions beyond stellar bodies include, for instance, the rigorous construction of thin-shell wormholes and the consistency of braneworld models in Palatini $f(R)$ gravity. Moreover, the approach followed here using distribution theory could be enlarged to include other Palatini theories of gravity beyond the $f(R)$ case, involving further contractions of the Ricci/Riemann tensors such as, for instance, in the class of the so-called Ricci-based gravities \cite{Afonso:2018bpv}, which are ghost-free, consistent extensions of GR. In these more general theories additional contributions of the stress-energy tensor (and not just its trace, as in the present case) are expected to yield novelties in the shape of the junction conditions as compared to GR, to Palatini $f(R)$ theories, and to their metric counterparts. This would allow, for instance, to reassess the claim of Ref.\cite{Pani:2012qd} on the existence of surface singularities in Eddington-inspired Born-Infeld gravity (see \cite{BeltranJimenez:2017doy} for a description of these theories, and \cite{Banados:2010ix,Tavakoli:2015llr} for specific applications) in pretty much the same way as for the Palatini $f(R)$ case studied here. Work along these lines is currently underway.

\section*{Acknowledgments}

 GJO is funded by the Ramon y Cajal contract RYC-2013-13019 (Spain). DRG is funded by the \emph{Atracci\'on de Talento Investigador} programme of the Comunidad de Madrid (Spain) No. 2018-T1/TIC-10431, and acknowledges further support from the Ministerio de Ciencia, Innovaci\'on y Universidades (Spain) project No. PID2019-108485GB-I00/AEI/10.13039/501100011033, and the Funda\c{c}\~ao para a Ci\^encia e a Tecnologia (FCT, Portugal) research projects Nos. PTDC/FIS-OUT/29048/2017 and PTDC/FIS-PAR/31938/2017.  This work is supported by the Spanish project  FIS2017-84440-C2-1-P (MINECO/FEDER, EU), the project H2020-MSCA-RISE-2017 Grant FunFiCO-777740, the project SEJI/2017/042 (Generalitat Valenciana), the Consolider Program CPANPHY-1205388, the Severo Ochoa grant SEV-2014-0398 (Spain)  and the Edital 006/2018 PRONEX (FAPESQ-PB/CNPQ, Brazil) Grant No. 0015/2019. This article is based upon work from COST Actions CA15117 and CA18108, supported by COST (European Cooperation in Science and Technology).

\end{document}